\documentclass{INTERSPEECH2023}

\interspeechcameraready

\usepackage[T1]{fontenc}
\usepackage{amsmath,graphicx}
\usepackage{epstopdf}
\usepackage{array}
\usepackage{graphicx}
\usepackage{multirow}
\usepackage{color, soul}
\usepackage{cite}
\usepackage{url}
\usepackage{threeparttable}
\usepackage{graphicx}
\usepackage{graphicx}
\usepackage{amsmath}
\usepackage{amssymb}

\title{Betray Oneself: A Novel Audio DeepFake Detection Model via \\Mono-to-Stereo Conversion}
\name{Rui Liu$^1$\thanks{This research is funded by the High-level Talents Introduction Project of Inner Mongolia University (No. 10000-22311201/002), the National Natural Science Foundation of China (NSFC) (No. 62206136, 62271432), and Agency for Science, Technology and Research (A*STAR) under its AME Programmatic Funding Scheme (Project No. A18A2b0046).}, Jinhua Zhang$^1$, Guanglai Gao$^1$ and Haizhou Li$^{2,3}$}
\address{
  $^1$Inner Mongolia University, China\\
  $^2$Shenzhen Research Institute of Big Data, School of Data Science,\\ The Chinese University of Hong Kong, Shenzhen, China
 \\
  $^3$ National University of Singapore, Singapore }
\email{liurui\_imu@163.com, zjh\_imu@163.com, csggl@imu.edu.cn, haizhouli@cuhk.edu.cn}

\begin{document}

\maketitle
 
\begin{abstract}
Audio Deepfake Detection (ADD) aims to detect the fake audio generated by text-to-speech (TTS), voice conversion (VC) and replay, etc., which is an emerging topic. Traditionally we take the mono signal as input and focus on robust feature extraction and effective classifier design. However, the dual-channel stereo information in the audio signal also includes important cues for deepfake, which has not been studied in the prior work. In this paper, we propose a novel ADD model, termed as M2S-ADD, that attempts to discover audio authenticity cues during the mono-to-stereo conversion process. We first projects the mono to a stereo signal using a pretrained stereo synthesizer, then employs a dual-branch neural architecture to process the left and right channel signals, respectively. In this way, we effectively reveal the artifacts in the fake audio, thus improve the ADD performance. The experiments on the ASVspoof2019 database show that M2S-ADD outperforms all baselines that input mono. We release the source code at \url{https://github.com/AI-S2-Lab/M2S-ADD}.
\end{abstract}

\section{Introduction}




%
The Deepfake technology seeks to adopt deep learning technology to produce audio that people have never said,  things that people have never done, and audio-visual content that has never existed
. Examples of its applications include face-shifting technology, voice simulation, face synthesis and
video generation~\cite{Ho2022VideoDM}. However, such technology can be abused. More and more deepfake content has been used for malicious purposes, such as in the spreading of fake news
and fraud cases~\cite{Mallet2023DeepfakeDA}. This calls for effective detection of the false content.

With the rapid progress of Text-To-Speech (TTS) \cite{9767637,9420276,liu2021GraphSpeech} synthesis and Voice Conversion (VC)
techniques, one is able to impersonate another's voice easily. The Audio Deepfake Detection (ADD) technology \cite{yi2022add} is a task that determines whether the given audio is authentic or counterfeited, and it has attracted increasing attention recently. Note that conventional methods for ADD mainly focus on two directions
, including 1) robust feature extraction and 2) effective model design. For the first direction,
Gupta et al. \cite{gupta2023replay} proposed a novel Cochlear Filter Cepstral Coefficients-based Instantaneous Frequency using Quadrature Energy Separation Algorithm (CFCCIF-QESA) features, with excellent temporal resolution as well as relative phase information. Some recent works
ingest raw audio~\cite{tak2021end} instead of hand-crafted acoustic feature and achieve robust ADD performance. For the second direction, researchers adopt Gaussian mixture model \cite{chettri2018deeper}, convolution neural network \cite{Lei2020}, deep neural networks \cite{9023289}, recurrent neural networks \cite{chen2018recurrent} etc. to build the ADD architecture. Recently, the graph neural network \cite{shim2022graph} was proposed to learn the relationships between cues at temporal intervals.


All of the above work dealt with single channel (mono) speech, and shown that ADD is clearly achievable. We note that two types of audio signals, namely single and dual-channel stereo, can be converted to each other~\cite{richard2021neural}.
The stereo signals provide a unique perspective on the speech quality of the audio signal~\cite{Delgado2019ObjectiveAO}, which has not been studied in the prior work of ADD. For example,
Tolooshams et al.~\cite{Tolooshams2021ATF} proposed a novel stereo-perception framework for speech enhancement. Compared with the traditional method with mono audio, using stereo can better preserve spatial images and enhance stereo mixture. Therefore, whether we can improve ADD performance by incorporating stereo information is the main concern of our work.
Fortunately, we found some inspiration from the field of computer vision (CV), that is, a face recognition system aims to confirm the identity of a person by their face~\cite{2020The}.
When performing face recognition, it is difficult to recognize the authenticity just rely on the realistic enough front face image \cite{2020Deep1}. However, the fake flaws will be exposed after rotating the face at a certain angle~\cite{2020Face}.

\begin{figure*}[h]
\includegraphics[width=\linewidth]{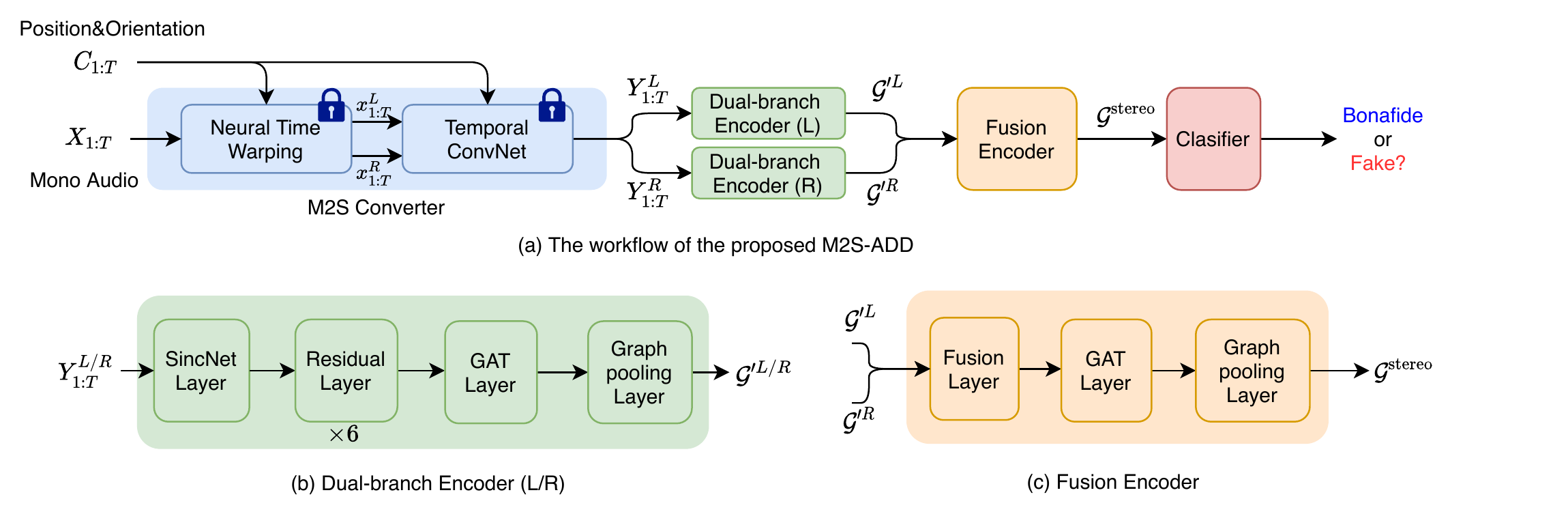}
\vspace{-7mm}
  \caption{The proposed M2S-ADD model (a)
 that consists of M2S Converter, Dual-branch Encoder (b), Fusion Encoder (c), and Classifier.
 The blue locks mean the parameter of those modules are initialized by pretraining and fixed during M2S-ADD training.}
 \vspace{-2mm}
  \label{fig:model}
\end{figure*}

Inspired by the above research, we propose a novel ADD model, termed as M2S-ADD, that attempts to discover audio authenticity cues during the mono-to-stereo (M2S) conversion process (similar to face rotation operation). Specifically, the M2S-ADD model first converts the audio of the mono channel into stereo through a pre-trained M2S conversion model and then employs a dual-branch neural architecture to process the left and right channel signals, respectively. Finally, the information from dual channels is fused for the final decision. By incorporating two-channel stereo information compared to mono speech, the fake audio could easily reveal itself and be successfully detected by our model.

Experiments on the mainstream ASVspoof 2019 logical access (LA) database show that our M2S-ADD model with stereo speech outperforms all baselines with mono speech. 
The main contributions of this paper include: 1) We propose a novel audio deepfake detection model, termed as M2S-ADD, via mono-to-stereo conversion; 2) We design a dual-branch neural architecture to learn the correlation between the two-channel audio signals and conduct joint training to optimize our M2S-ADD model; 
To our best knowledge, this is the first study of audio deepfake detection with stereo speech data augmentation. 

\section{M2S-ADD Neural Architecture}
\label{m2sADD}
\vspace{-1mm}
We propose a neural architecture, termed as M2S-ADD, as shown in Fig.~\ref{fig:model}, that consists of M2S converter, dual-branch encoder, fusion encoder, and classifier.
The M2S converter converts the original mono audio into high-quality stereo, which includes two channel signals. The dual-branch encoder reads the two-channel audio and extracts the high-level feature representations for the left and right channels separately. The fusion encoder combines the two-channel feature representations together to facilitate the modeling of information between different dimensions. At last, the two-class classifier is to detect the authenticity (bonafide or fake). As M2S-ADD is expected to be better informed than the mono input system, it greatly reduces the model complexity.

\vspace{-2mm}
\subsection{M2S Converter}
\vspace{-2mm}

The M2S converter aims to transform the single-channel mono signal into dual-channel stereo signal. As shown in the blue panel of Fig.\ref{fig:model}(a), the M2S converter consists of neural time warping and temporal ConvNet. The neural time warping module learns an accurate warp from the source position to the listener's left and right ear while respecting physical properties like monotonicity and causality. The Temporal ConvNet module nuanced effects like room reverberations or head- and ear-shape related modifications to the signal.

Specifically, given a monaural (single-channel) signal $X_{1:T} = (X_{1}, X_{2},...,X_{T})$ of length $T$ and its conditioning temporal signal $C_{1:T} = (C_{1}, C_{2},..., C_{T})$ that represents the position and orientation of source and listener respectively, neural time warping first adopts a CNN based WarpNet to read $C_{1:T}$ to predict a neural warpfield $\rho'$ and add it to the herometric warpfield $\rho$. Then, the final warpfield $\rho + \rho'$ was used to compute the two-channel warped signal $x^{L/R}_{1:T}$, where the channels represent left and right ear, by a recursive activation function. At last, the temporal ConvNet includes WaveNet like conditional temporal convolutions to input the $x^{L/R}_{1:T}$ and $C_{1:T}$ to reconstruct the expected dual-channel stereo signal $Y^{L/R}_{1:T}$.


The M2S converter share same architecture with the latest neural sound binauralization model \cite{richard2021neural}, which includes more details. It's worth mentioning that we adopt such a pure end-to-end network as part of our M2S-ADD model to facilitate the entire pipeline. The parameters of the M2S converter need to be pre-trained and then fixed during M2S-ADD training to achieve a stable M2S conversion, as shown in the blue locks of Fig.\ref{fig:model}(a).


\subsection{Dual-branch Encoder}
Dual-branch encoder is devised to extract high-level and meaningful feature representation for dual-channel stereo signals. As shown in Fig.\ref{fig:model}(a) and (b), Dual-branch encoder (L) and Dual-branch encoder (R) share the same structure to process left and right channel data respectively. The dual-branch encoder is composed by one SincNet layer, six Residual layers, one graph attention network (GAT) layer and one Graph pooling layer. Note that one SincNet layer and six Residual layers were stacked to let the CNN based network deepen and introduce more nonlinearity to learn the detail clues in the left and right channels audio signal.
The GAT mechanism has been proven to learn the relationships between artifacts in different sub-bands or temporal intervals \cite{202113}. Following this spirit, the GAT layer is used to aggregate the associated information by using the self-attention weight between information pairs and the graph pooling layer used to discard useless and duplicate information.

Specifically, the SincNet layer consists of Conv-1D layer, max-pooling layer, Batch Normalization (BN) and SeLU activation function.
The residual layer consists of a Conv-2D layer with BN and SeLU and another Conv-2D layer with max-pooling. SincNet and residual layers of dual-branch encoder (L) or (R) read the left or right signal to extract high-level feature respectively:
\vspace{-1mm}
 \begin{equation}
  \begin{aligned}
     &h'^{L}_{1:T} = {f^L}(Y^{L}_{1:T})\\
    &h'^{R}_{1:T} = {f^R}(Y^{R}_{1:T})
  \end{aligned}
 \end{equation}
where ${f}$ means the CNN network was stacked by the SincNet and the residual layers.

The GAT layer operates upon an input graph $\mathcal{G}$ to produce an output graph $\mathcal{G'}$. Note that $\mathcal{G}$ is formed from the high-level feature $h'^{L/R}_{1:T}$, and defined as $\mathcal{G}(N, \varepsilon, h')$.
$N$ is the set of nodes, also the number of temporal frames after spectral averaging. $\varepsilon$ represents the edges between all possible node connections, including self-connections.

Then, GAT aggregates neighboring nodes using learnable weights via a self-attention mechanism.
The attention weight $\alpha_{u,n}$ between nodes $u$ and $n$ are calculated according to:
\vspace{-1.6mm}
\begin{equation}
\alpha_{u, n}=\frac{\exp \left(W_{\left(h^{\prime}{ }_n \odot h^{\prime} u\right)}\right)}{\sum_{w \in \mathcal{M}(n) \cup\{n\}} \exp \left(W\left(h^{\prime}{ }_n \odot h^{\prime}{ }_w\right)\right)}
\end{equation}
where $\mathcal{M}(n)$ refers to the set of neighboring nodes for the node $n$. $W$ is the learnable weight and $\odot$ means element-level multiplication.

The output graph $\mathcal{G'}$ comprises a set of nodes $o$, where each node $o$ is computed by:
\vspace{-2mm}
\begin{equation}
\begin{aligned}
&\mathrm{o}_n=\operatorname{SeLU}\left(\mathrm{BN}\left(  m_n +  h'_n \right)  \right) \\
&m_n = \sum_{u \in \mathcal{M}(n) \cup\{n\}} \alpha_{u, n} h'_u
\end{aligned}
\vspace{-1mm}
\end{equation}

Graph pooling generates more discriminative graphs by selecting a subset of the most informative nodes. We follow \cite{gao2019graph} and generate the pooled graph representations $\mathcal{G}'^{L}$ and $\mathcal{G}'^{R}$ from the original $\mathcal{G}^{L}$ and $\mathcal{G}^{R}$ graphs of left and right channels, respectively. At last, both graph nodes are projected into the same dimensional space using an affine-transform to match the input node dimensionality of the fusion encoder.

\vspace{-2mm}
\subsection{Fusion Encoder and Classifier}
\vspace{-1mm}
The fusion encoder is used to exploit complementary information captured by the attention graphs of left and right channels.
As shown in Fig.\ref{fig:model}(c), the fusion encoder includes fusion layer, GAT layer and the graph pooling layer. The fusion layer acts to  combine $\mathcal{G}'^{L}$ and $\mathcal{G}'^{R}$ to a fused graph $\mathcal{G'}^{\mathrm{stereo}}$ using element-wise multiplication:
\vspace{-2mm}
\begin{equation}
    \mathcal{G'}^{\mathrm{stereo}}=  \mathcal{G'}^{L} \odot \mathcal{G'}^{R}.
    \vspace{-2mm}
\end{equation}

The GAT layer is applied to  $\mathcal{G'}^{\mathrm{stereo}}$ and the graph pooling is then used to generate a pooled graph $\mathcal{G}^{\mathrm{stereo}}$.  Unlike the GAT and graph pooling layers in dual-branch encoder were processed for left and right channel graphs, the GAT  and graph pooling layers in this section performs further feature extraction on the fused stereo graph, which will further analyze useful cues for audio deepfake detection, avoiding the loss of useful information and the side effect of redundant information. At last, we use a fully-connected projection layer to obtain the final two-class prediction (bona fide or fake).


By carefully analyzing the left and right channel signals in two-channel stereo audio, our method finds more important fake audio cues than mono audio analysis methods, allowing for accurate fake audio detection.

\vspace{-1mm}
\section{Experiments}
\label{sec:exp}
\vspace{-1mm}

\subsection{Dataset}
\textbf{M2S Converter Pretraining}: An $<$mono, stereo$>$ paired audio data with appropriately size is a prerequisite for M2S converter pretraining. We follow ~\cite{richard2021neural} and conduct pretraining on a total of 2 hours of paired mono and binaural data at 48kHz from eight different speakers, four male and four female. Specifically, a mannequin equipped with binaural microphones in its ears is treated as the listener. Participants were asked to walk around the mannequin in a circle with 1.5m radius and make a spontaneous conversation with it. The conditioning temporal signals $C_{1:T}$ for each audio are recorded from the realistic scenario \footnote{Pretraining dataset link: \url{https://github.com/facebookresearch/BinauralSpeechSynthesis/releases/download/v1.0/binaural_dataset.zip}} \cite{richard2021neural}.
More detailed descriptions are referred to~\cite{richard2021neural}.

\noindent \textbf{M2S-ADD Training}: We use the official ASVspoof 2019 logical access (LA) database\footnote{\url{https://www.asvspoof.org/index2019.html}} to validate the performance of M2S-ADD. The real audio in the LA database is directly selected from the VCTK corpus, while the fake audio in the database is obtained by using different TTS and VC systems.
The sampling rate of all audios is 16kHz. The data in the database is divided into three subsets, training set, development set and evaluation set. Among them, the fake audio in the training and development set all come from 6 systems (A01-A06), while the fake audio in the evaluation set was created with 13 systems (A07-A19).

\begin{table}[]
\caption{The configuration for our M2S-ADD model.}
\vspace{-3mm}
\label{tab:para}
\resizebox{\linewidth}{!}{  
    \begin{tabular}{lll} %
        \toprule 
        \textbf{Layer} & \textbf{Input:64600 samples} & \textbf{Output shape} \\
        \midrule 
              & \textbf{Dual-branch Encoder(L/R)} & \\
        \midrule 
          SincNet Layer & Conv-1D(129,1,70) & (70,64472)\\
                        & Maxpool-2D(3)  & (1,23,21490)\\
                        & BN\&SeLU & \\
        \cline{2-3}
        Residual Layer & $\left\{\begin{array}{c}
\text { Conv$\mbox{-}$2D ((2,3), 1,32)} \\
\text { BN \& SeLU } \\
\text { Conv$\mbox{-}$2D ((2,3), 1,32)} \\
\text { Maxpool$\mbox{-}$2D ((1,3))}
\end{array}\right\} \times$ 2 &(32,23,2387)\\

             & $\left\{\begin{array}{c}
\text { Conv$\mbox{-}$2D  ((2,3), 1,64)} \\
\text { BN \& SeLU } \\
\text { Conv$\mbox{-}$2D  ((2,3), 1,64)} \\
\text { Maxpool$\mbox{-}$2D  ((1,3))}
\end{array}\right\} \times$ 4 &(64,23,29)   \\
        \cline{2-3}

        GAT Layer & GAT Layer & $\begin{array}{c}  left:(32,23)  \\  right:(32,29)\end{array}$\\
        \cline{2-3}
        Graph Pooling Layer(1) & Graph pooling & $\begin{array}{c}  left:(32,14)  \\  right:(32,23)\end{array}$ \\
            & Projection & (32,12)\\
        \midrule
            & \textbf{Fusion Encoder} & \\
        \midrule
        Fusion Layer & Element-wise multiplication & (32,12) \\
        \cline{2-3}
        GAT Layer & GAT Layer & (16,12) \\
        \cline{2-3}
        Graph Pooling Layer(2) & Graph Pooling &(16,7) \\
                               & Projection &(1,7) \\
        \midrule
            & \textbf{Classifier} & \\
        \midrule
        Classifier & FC(2) & 2 \\

        \bottomrule 
    \end{tabular}
}
\vspace{-7mm}
\end{table}

\subsection{Experimental Setup}
{For M2S converter pretraining, we train it for 100 epochs using an Adam optimizer.
It's worth mentioning that the original M2S converter takes a mono audio segment with a length of 9600 samples as input and then outputs the left and right channels with the same length. To achieve utterance-level mono-to-stereo conversion for ASVspoof 2019 LA database during M2S converter, we first split each audio into various segments with 64600 samples as the input, then merge the output to obtain the utterance-level left and right channels. Note that the conditioning temporal signals $C_{1:T}$ for each audio during mono-to-stereo conversion are randomly sampled from real data$^{1}$.
For M2S-ADD model, the Dual-branch Encoder(L/R) consists of SincNet Layer, 6 Residual Layers, GAT Layer, and a Graph pooling Layer. In SincNet Layer, the input length of data is 64600 samples, and the number of filters is 70. In the Residual Layer, we use 32 and 64 filters in Conv-2D of two kinds of residual blocks respectively to further improve the generalization ability of the model to cope with unknown attacks. In the GAT Layer, we use absolute values to prevent useful information with negative values from being discarded. A more detailed summary of the configuration for our M2S-ADD model is shown in Table \ref{tab:para}.

In order to ensure that the results of each training are consistent, we use the random number seed to initialize parameters, whose default value is 1234. The model has trained for 400 epochs, the batch size is 24; the learning rate is 0.0001; the weight decay rate is 0.0001; the loss function is Weighted Cross Entropy Loss.
Our program runs in the environment of python 3.8, and the driver Version is 515.48.07, and the CUDA version is 11.7. We report the results in terms of Equal Error Rate (EER)~\cite{9143410} rather than the minimum normalized tandem detection cost function (min-tDCF) ~\cite{kinnunen18b_odyssey} since min-tDCF is more applicable to speaker verification task~\cite{9143410}.}

\begin{table}[]\footnotesize
  \caption{Comparison of M2S-ADD with all baselines with mono input.
  ($^{*}$ means we retrained this graph-based model using our own GPU and hyperparameters as MS-ADD, like batch size, for a fair comparison and achieved different results from the original paper.)
  }
  \vspace{-1mm}
  \label{tab:comparison}
  \centering
  \begin{threeparttable}
  \begin{tabular}{p{4.3cm}<{\centering}p{2.5cm}<{\centering}}
    \toprule
    \textbf{System}      & \textbf{EER} ($\downarrow$)               \\
    \midrule
    ResNet~\cite{202164}                & 3.72        \\
    LCNN~\cite{201962}                  & 5.06         \\
    LCNN-4CBAM~\cite{202158}            & 3.67         \\
    Siamese CNN~\cite{Lei2020}            & 3.79         \\
    PC-DARTS~\cite{202159}               & 4.96           \\
    ResNet18-GAT-S~\cite{202113}           & 4.48         \\
    GMM~\cite{tak2020spoofing}          & 3.50           \\
    ResNet18-GAT-T~\cite{202113}         & 4.71         \\
    Resnet18-AM-Softmax~\cite{202156}     & 3.26         \\
    LCNN-Dual attention~\cite{202158}     & 2.76         \\
    SE-Res2Net50~\cite{202157}      & 2.50         \\
    MLCG-Res2Net50+CE~\cite{202154}     & 2.15        \\
    Resnet18-OC-softmax~\cite{202156}     & 2.19         \\
    Capsule network~\cite{202155}     & 1.97         \\
    LCNN-LSTM-sum~\cite{202153}          & 1.92            \\
    ResNet18-LMCL-FM~\cite{202014}       & 1.81           \\
    MCG-Res2Net50+CE~\cite{Li2021ChannelwiseGR}    & 1.78  \\
    Res-TSSDNet~\cite{9456037}        & 1.64               \\
    $^{*}$ RawGAT-ST
    ~\cite{2021End}        &1.39  \\
    \hline
     \textbf{M2S-ADD \emph{(proposed)}}      & \textbf{1.34} \\
     \hline
    \qquad w/o dual-branch            &2.46                   \\
    \bottomrule
  \end{tabular}
  \vspace{-4mm}
  \end{threeparttable}
\end{table}

\subsection{Comparative Study}

This work is one of the first attempts to exploit stereo information in ADD field. We choose some state-of-the-art ADD systems, with mono audio as an input, as the benchmark. Table~\ref{tab:comparison} illustrates a comparison of performance for the proposed M2S-ADD model and various baselines with mono input.


Note that the RawGAT-ST model is very similar to our model. For a fair comparison, we retrained it on our own GPU using the same hyperparameters as M2S-ADD, such as batch size. The results show that our model outperforms it in terms of EER. In a nutshell, the results show that our model outperforms all other models.


 \vspace{-2mm}
\subsection{Ablation Study}
 \vspace{-2mm}
To further validate our M2S-ADD model, we conduct an ablation experiment in this section. Specifically,
to verify the validity of the dual-branch neural architecture, we reconstructed the mono signal into a stereo and then averaged the information from the left and right channels globally and fed it to the dual-channel encoder (L) or (R) for feature extraction. The results are shown in the last row of Table~\ref{tab:comparison}. We can find that abandoning the dual-branch mechanism to directly encode information from the left and right channels would disrupt the modeling of information interaction between them, resulting in a significant reduction in ADD performance.

\begin{figure}[h]
\centering
\begin{minipage}{\linewidth}
  \centerline{
  \includegraphics[width=.88\linewidth]{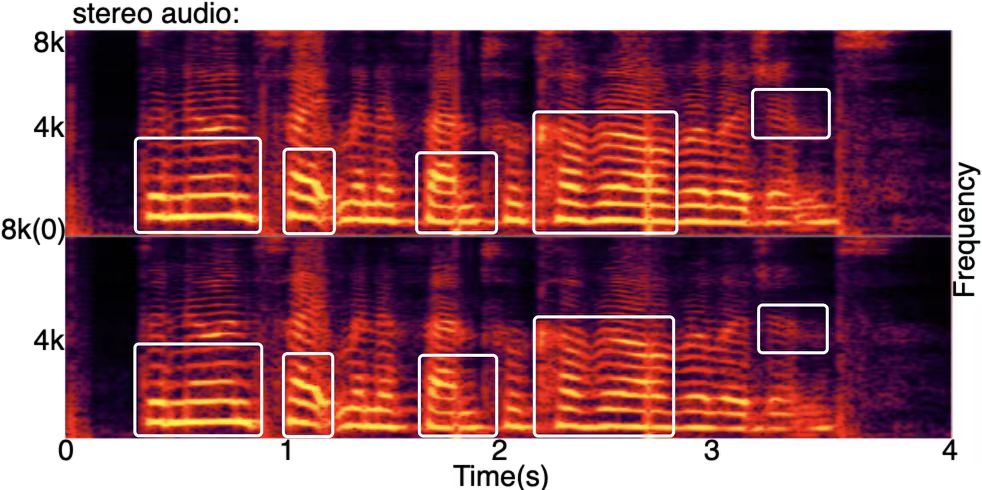}

  }
\end{minipage}
\hfill
\begin{minipage}{\linewidth}
  \centerline{
  \includegraphics[width=.88\linewidth]{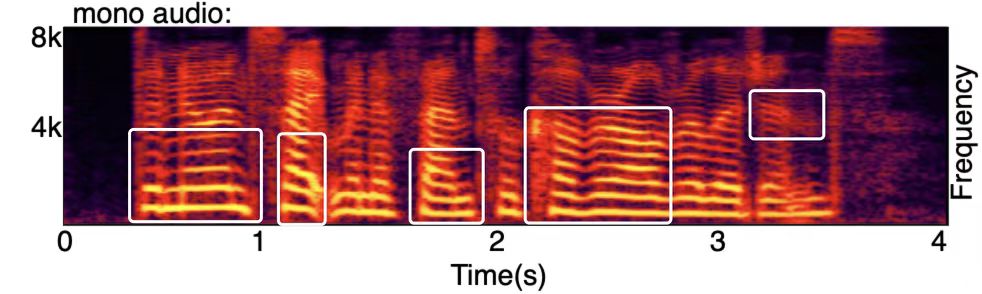}

  }
  \centerline{(a) Bonafide audio}
\end{minipage}
\vfill
\begin{minipage}{\linewidth}
   \centerline{
  \includegraphics[width=.88\linewidth]{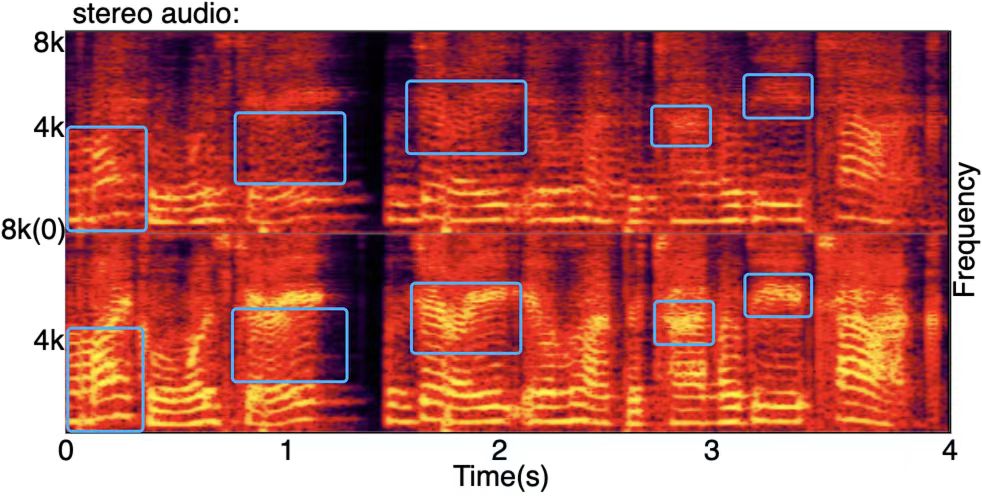}

  }
\end{minipage}
\hfill
\begin{minipage}{\linewidth}
   \centerline{
  \includegraphics[width=.88\linewidth]{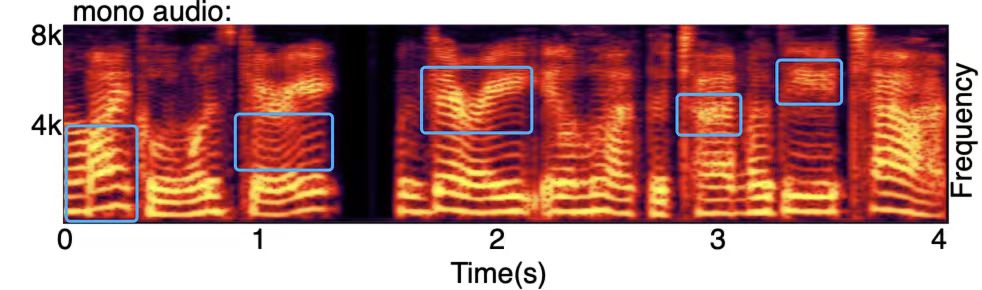}
  }
  \centerline{(b) Fake audio}
\end{minipage} 
 \vspace{-2mm}
\caption{Comparison of visualization analysis in the spectral domain. (a) shows the spectral details of the mono and stereo signals for bonafide audio; (b) shows the spectral details for fake audio. Unlike the white boxes, the blue boxes show that the spectral artifacts are particularly exposed when the mono fake audio is converted to stereo audio.}
\vspace{-3mm}
\label{fig:comparison}
\end{figure}

\vspace{-2mm}
\subsection{Visualization Study}
\vspace{-2mm}
In order to check the effectiveness of our method more visually, we have tried to analyze it in a visualization study. As a visualization of the time domain signal is not intuitive enough, we converted the time domain signal into spectral features and then observed the fine-grained differences between bonafide and fake audio. 

As shown in Fig.\ref{fig:comparison}(a) and (b), we list the spectral features for bonafide ad fake audios in terms of mono and stereo channels. White and blue boxes are used to highlight the spectral details at specific locations for bonafide ad fake audios respectively.
As can be seen from the blue boxes for fake audio, the original spectral information is not fully reflected, and there is even a lot of noise, in the left and right channels of the stereo panel after mono-to-stereo conversion. Note that all the spectral details for bonafide audio, as in the white boxes, are preserved.
In a nutshell, it was once again verified that converting mono audio to stereo audio can expose important cues among the fake audio, thus greatly reducing the difficulty of ADD.

    




\vspace{-2mm}
\section{Conclusions}
\label{sec:con}
\vspace{-2mm}
In this work, we propose a novel audio deepfake detection model, denoted as M2S-ADD. The M2S-ADD model can handle dual-channel stereo signals and adopts dual-branch architecture to learn the subtle information among the left and right channel signals, thus effectively discovering audio authenticity cues during the mono-to-stereo conversion process. Experimental results demonstrate that our M2S-ADD can achieve accurate ADD performance for various attacks with the help of authenticity cues from dual-channel stereo signals. As per our knowledge, the proposed M2S-ADD is the first end-to-end ADD model that by exploiting stereo information. In future work, we will continue to optimize the model structure and exploit the detailed spectral information to further boost the ADD performance.

\bibliographystyle{IEEEtran}
{\tiny
\bibliography{mybib}

\begin{thebibliography}{10}
\providecommand{\url}[1]{#1}
\csname url@samestyle\endcsname
\providecommand{\newblock}{\relax}
\providecommand{\bibinfo}[2]{#2}
\providecommand{\BIBentrySTDinterwordspacing}{\spaceskip=0pt\relax}
\providecommand{\BIBentryALTinterwordstretchfactor}{4}
\providecommand{\BIBentryALTinterwordspacing}{\spaceskip=\fontdimen2\font plus
\BIBentryALTinterwordstretchfactor\fontdimen3\font minus
  \fontdimen4\font\relax}
\providecommand{\BIBforeignlanguage}[2]{{%
\expandafter\ifx\csname l@#1\endcsname\relax
\typeout{** WARNING: IEEEtran.bst: No hyphenation pattern has been}%
\typeout{** loaded for the language `#1'. Using the pattern for}%
\typeout{** the default language instead.}%
\else
\language=\csname l@#1\endcsname
\fi
#2}}
\providecommand{\BIBdecl}{\relax}
\BIBdecl

\bibitem{Ho2022VideoDM}
J.~Ho, T.~Salimans, A.~Gritsenko, W.~Chan, M.~Norouzi, and D.~J. Fleet, ``Video
  diffusion models,'' \emph{ArXiv}, vol. abs/2204.03458, 2022.

\bibitem{Mallet2023DeepfakeDA}
J.~Mallet, L.~Pryor, R.~Dave, and M.~Vanamala, ``Deepfake detection analyzing
  hybrid dataset utilizing cnn and svm,'' \emph{ArXiv}, vol. abs/2302.10280,
  2023.

\bibitem{9767637}
R.~Liu, B.~Sisman, G.~l. Gao, and H.~Li, ``Decoding knowledge transfer for
  neural text-to-speech training,'' \emph{IEEE/ACM Transactions on Audio,
  Speech, and Language Processing}, pp. 1--1, 2022.

\bibitem{9420276}
R.~Liu, B.~Sisman, G.~Gao, and H.~Li, ``Expressive tts training with frame and
  style reconstruction loss,'' \emph{IEEE/ACM Transactions on Audio, Speech,
  and Language Processing}, vol.~29, pp. 1806--1818, 2021.

\bibitem{liu2021GraphSpeech}
R.~Liu, B.~Sisman, and H.~Li, ``Graphspeech: Syntax-aware graph attention
  network for neural speech synthesis,'' in \emph{ICASSP 2021 - 2021 IEEE
  International Conference on Acoustics, Speech and Signal Processing
  (ICASSP)}, 2021, pp. 6059--6063.

\bibitem{yi2022add}
J.~Yi, R.~Fu, J.~Tao, S.~Nie, H.~Ma, C.~Wang, T.~Wang, Z.~Tian, Y.~Bai, C.~Fan
  \emph{et~al.}, ``Add 2022: the first audio deep synthesis detection
  challenge,'' in \emph{ICASSP 2022-2022 IEEE International Conference on
  Acoustics, Speech and Signal Processing (ICASSP)}.\hskip 1em plus 0.5em minus
  0.4em\relax IEEE, 2022, pp. 9216--9220.

\bibitem{gupta2023replay}
P.~Gupta, P.~K. Chodingala, and H.~A. Patil, ``Replay spoof detection using
  energy separation based instantaneous frequency estimation from quadrature
  and in-phase components,'' \emph{Computer Speech \& Language}, vol.~77, p.
  101423, 2023.

\bibitem{tak2021end}
H.~Tak, J.~Patino, M.~Todisco, A.~Nautsch, N.~Evans, and A.~Larcher,
  ``End-to-end anti-spoofing with rawnet2,'' in \emph{ICASSP 2021-2021 IEEE
  International Conference on Acoustics, Speech and Signal Processing
  (ICASSP)}.\hskip 1em plus 0.5em minus 0.4em\relax IEEE, 2021, pp. 6369--6373.

\bibitem{chettri2018deeper}
B.~Chettri and B.~L. Sturm, ``A deeper look at gaussian mixture model based
  anti-spoofing systems,'' in \emph{2018 IEEE International Conference on
  Acoustics, Speech and Signal Processing (ICASSP)}.\hskip 1em plus 0.5em minus
  0.4em\relax IEEE, 2018, pp. 5159--5163.

\bibitem{Lei2020}
Z.~Lei, Y.~Yang, C.~Liu, and J.~Ye, ``Siamese convolutional neural network
  using gaussian probability feature for spoofing speech detection.''
  \emph{Proc. Interspeech 2020}, pp. 1116--1120, 2020.

\bibitem{9023289}
J.~Li, M.~Sun, and X.~Zhang, ``Multi-task learning of deep neural networks for
  joint automatic speaker verification and spoofing detection,'' in \emph{2019
  Asia-Pacific Signal and Information Processing Association Annual Summit and
  Conference (APSIPA ASC)}, 2019, pp. 1517--1522.

\bibitem{chen2018recurrent}
Z.~Chen, W.~Zhang, Z.~Xie, X.~Xu, and D.~Chen, ``Recurrent neural networks for
  automatic replay spoofing attack detection,'' in \emph{2018 IEEE
  international conference on acoustics, speech and signal processing
  (ICASSP)}.\hskip 1em plus 0.5em minus 0.4em\relax IEEE, 2018, pp. 2052--2056.

\bibitem{shim2022graph}
H.-j. Shim, J.~Heo, J.-h. Park, G.-h. Lee, and H.-J. Yu, ``Graph attentive
  feature aggregation for text-independent speaker verification,'' in
  \emph{ICASSP 2022-2022 IEEE International Conference on Acoustics, Speech and
  Signal Processing (ICASSP)}.\hskip 1em plus 0.5em minus 0.4em\relax IEEE,
  2022, pp. 7972--7976.

\bibitem{richard2021neural}
A.~Richard, D.~Markovic, I.~D. Gebru, S.~Krenn, G.~A. Butler, F.~Torre, and
  Y.~Sheikh, ``Neural synthesis of binaural speech from mono audio,'' in
  \emph{International Conference on Learning Representations}, 2021.

\bibitem{Delgado2019ObjectiveAO}
P.~M. Delgado and J.~Herre, ``Objective assessment of spatial audio quality
  using directional loudness maps,'' \emph{ICASSP 2019 - 2019 IEEE
  International Conference on Acoustics, Speech and Signal Processing
  (ICASSP)}, pp. 621--625, 2019.

\bibitem{Tolooshams2021ATF}
B.~Tolooshams and K.~Koishida, ``A training framework for stereo-aware speech
  enhancement using deep neural networks,'' \emph{ICASSP 2022 - 2022 IEEE
  International Conference on Acoustics, Speech and Signal Processing
  (ICASSP)}, pp. 6962--6966, 2021.

\bibitem{2020The}
H.~Du, H.~Shi, D.~Zeng, and T.~Mei, ``The elements of end-to-end deep face
  recognition: A survey of recent advances,'' 2020.

\bibitem{2020Deep1}
M.~Abdolahnejad and P.~X. Liu, ``Deep learning for face image synthesis and
  semantic manipulations: a review and future perspectives,'' \emph{Artificial
  Intelligence Review}, vol.~53, no.~3, 2020.

\bibitem{2020Face}
E.~Winarno, I.~Amin, S.~Hartati, and P.~W. Adi, ``Face recognition based on cnn
  2d-3d reconstruction using shape and texture vectors combining,''
  \emph{Indonesian Journal of Electrical Engineering and Informatics (IJEEI)},
  no.~2, 2020.

\bibitem{202113}
J.~P. M.~T. H.~Tak, J.-w.~Jung and N.~Evans, ``Graph attention networks for
  anti-spoofing,'' \emph{Proc. INTERSPEECH}, 2021.

\bibitem{gao2019graph}
H.~Gao and S.~Ji, ``Graph u-nets,'' in \emph{international conference on
  machine learning}.\hskip 1em plus 0.5em minus 0.4em\relax PMLR, 2019, pp.
  2083--2092.

\bibitem{9143410}
T.~Kinnunen, H.~Delgado, N.~Evans, K.~A. Lee, V.~Vestman, A.~Nautsch,
  M.~Todisco, X.~Wang, M.~Sahidullah, J.~Yamagishi, and D.~A. Reynolds,
  ``Tandem assessment of spoofing countermeasures and automatic speaker
  verification: Fundamentals,'' \emph{IEEE/ACM Transactions on Audio, Speech,
  and Language Processing}, vol.~28, pp. 2195--2210, 2020.

\bibitem{kinnunen18b_odyssey}
T.~Kinnunen, K.~A. Lee, H.~Delgado, N.~Evans, M.~Todisco, M.~Sahidullah,
  J.~Yamagishi, and D.~A. Reynolds, ``{t-DCF: a Detection Cost Function for the
  Tandem Assessment of Spoofing Countermeasures and Automatic Speaker
  Verification },'' in \emph{Proc. The Speaker and Language Recognition
  Workshop (Odyssey 2018)}, 2018, pp. 312--319.

\bibitem{202164}
R.~K.~D. J.~Yang, H.~Wang and Y.~Qian, ``Modified magnitude-phase spectrum
  information for spoofing de- tection,'' \emph{IEEE/ACM TASLP}, 2021.

\bibitem{201962}
G.~Lavrentyeva, S.~Novoselov, A.~Tseren, M.~Volkova, A.~Gorlanov, and
  A.~Kozlov, ``Stc antispoofing systems for the asvspoof2019 challenge,''
  \emph{Proc. Interspeech 2019}, pp. 1033--1037, 2019.

\bibitem{202158}
S.~Z. S.~H. X.~Ma, T.~Liang and L.~He, ``Improved lightcnn with attention
  modules for asv spoofing detection,'' \emph{2021 IEEE International
  Conference on Multimedia and Expo (ICME)}, vol. pp. 1–6, 2021.

\bibitem{202159}
W.~Ge, M.~Panariello, J.~Patino, M.~Todisco, and N.~Evans,
  ``Partially-connected differentiable architecture search for deepfake and
  spoofing detection,'' in \emph{Interspeech 2021}.\hskip 1em plus 0.5em minus
  0.4em\relax ISCA, 2021, pp. 4319--4323.

\bibitem{tak2020spoofing}
H.~Tak, J.~Patino, A.~Nautsch, N.~Evans, and M.~Todisco, ``Spoofing attack
  detection using the non-linear fusion of sub-band classifiers,'' \emph{Proc.
  Interspeech 2020}, pp. 1106--1110, 2020.

\bibitem{202156}
F.~J. Y.~Zhang and Z.~Duan, ``One-class learning to- wards synthetic voice
  spoofing detection,'' \emph{IEEE Signal Processing Letters}, vol. pp.
  937–941, 2021.

\bibitem{202157}
C.~X. D. D.~a. X.Li, N.Li, ``Replay and synthetic speech detection with res2net
  architecture,'' \emph{Proc. ICASSP}, vol. pp. 6354–6358, 2021.

\bibitem{202154}
H.~L. X.~L. X.~Li, X.~Wu and H.~Meng, ``Channel- wise gated res2net: Towards
  robust detection of synthetic speech attacks,'' \emph{Proc. INTERSPEECH},
  2021.

\bibitem{202155}
Y.~X.~a. A.Luo, E.Li, ``Acapsule network based approach for detection of audio
  spoofing attacks,'' \emph{Proc. ICASSP}, vol. pp. 6359–6363, 2021.

\bibitem{202153}
X.~Wang and J.~Yamagishi, ``A comparative study on re- cent neural spoofing
  countermeasures for synthetic speech detection,'' \emph{Proc. INTERSPEECH},
  2021.

\bibitem{202014}
P.~N. G.~S. T.~Chen, A.~Kumar and E.~Khoury, ``Generalization of audio deepfake
  detection,'' \emph{Proc. Speaker Odyssey Workshop}, vol. pp. 132–137, 2020.

\bibitem{Li2021ChannelwiseGR}
X.~Li, X.~Wu, H.~Lu, X.~Liu, and H.~M. Meng, ``Channel-wise gated res2net:
  Towards robust detection of synthetic speech attacks,'' \emph{ArXiv}, vol.
  abs/2107.08803, 2021.

\bibitem{9456037}
G.~Hua, A.~B.~J. Teoh, and H.~Zhang, ``Towards end-to-end synthetic speech
  detection,'' \emph{IEEE Signal Processing Letters}, vol.~28, pp. 1265--1269,
  2021.

\bibitem{2021End}
H.~Tak, J.~W. Jung, J.~Patino, M.~Kamble, M.~Todisco, and N.~Evans,
  ``End-to-end spectro-temporal graph attention networks for speaker
  verification anti-spoofing and speech deepfake detection,'' 2021.

\end{thebibliography}
}

\end{document}